\documentclass[epsfig,12pt,onecolumn]{article}
\usepackage{amssymb}
\usepackage{amsmath}
\usepackage{multicol}
\usepackage{graphicx}
\usepackage{float}
\usepackage{caption}

\restylefloat{table}
\graphicspath{ {./images/} }
\input{tcilatex}
\makeatletter \@addtoreset{equation}{section}

\def \be{\begin{equation}}
\def \ee{\end{equation}}
\def \bea{\begin{eqnarray}}
\def \eea{\end{eqnarray}}

\newcommand{\nc}{\newcommand}
\nc{\al}{\alpha} \nc{\bib}{\bibitem} \nc{\la}{\lambda}
\nc{\C}{\mbox{\hspace{1.24mm}\rule{0.2mm}{2.5mm}\hspace{-2.7mm} C}}
\nc{\R}{\mbox{\hspace{.04mm}\rule{0.2mm}{2.8mm}\hspace{-1.5mm} R}}

\begin{document}

\title{Charged 4D Einstein-Gauss-Bonnet Black Hole: Vacuum solutions, Cauchy
Horizon, Thermodynamics}
\author{M. Bousder$^{1}$\thanks{%
mostafa.bousder@um5.ac.ma}, K. El Bourakadi$^{2}$\thanks{%
k.elbourakadi@yahoo.com} and M.Bennai$^{1,2}$\thanks{%
mohamed.bennai@univh2c.ma} \\
$^{1}${\small LPHE-MS Laboratory, Department of physics,}\\
\ {\small Faculty of Science, Mohammed V University in Rabat, Rabat, Morocco}%
\\
$^{2}${\small Physics and Quantum Technology team, LPMC, Ben Mâ€%
™sik Faculty of Sciences,}\\
{\small Hassan II de Casablanca University, Morocco}}
\maketitle

\begin{abstract}
In this paper, we investigate the four-dimensional Einstein-Gauss-Bonnet
black hole. The thermodynamic variables and equations of state of black
holes are obtained in terms of a new parameterization. We discuss a
formulation of the van der Waals equation by studying the effects of the
temperature on $P-V$ isotherms. We show the influence of the Cauchy horizon
on the thermodynamic parameters. We prove by different methods, that the
black hole entropy obey area law (plus logarithmic term that\textbf{\ }%
depends on the Gauss-Bonnet coupling $\alpha $). We propose a physical
meaning for the logarithmic correction to the area law. This work can be
extended to the extremal EGB black hole, in that case, we study the
relationship between compressibility factor, specific heat and the coupling $%
\alpha $.

\textbf{Keywords:} Black hole, Entropy, Einstein-Gauss-Bonnet gravity, van
der Waals fluid.
\end{abstract}

\section{Introduction}

The study of Einstein-Gauss-Bonnet gravity (EGB) is very important as it
offers a more comprehensive set up to explore many gravity-related
conceptual problems. EGB gravity is a major, higher dimensional
generalization of Einstein gravity. Lovelock's theorem \cite{G3}, suggests
that Einstein's general relativity is a theory of gravity that respects
various aspects such as spacetime is $4-$dimensional. Recently, Glavan and
Lin in Ref. \cite{G1}, introduced a general covariant modified theory of
gravity in $4-$spacetime dimensions which propagates only the massless
graviton and also bypasses Lovelock's theorem. The case of $4-$dimensional
EGB gravity is noteworthy because the Euler-Gauss Bonnet term becomes a
topological invariant, whereby the equations of motion and the gravitational
dynamics are not affected. The intriguing idea of D. Glavan and C. Lin was
to multiply the GB term by the factor $1/(D-4)$ before taking the limit $%
D\rightarrow 4$. This technique offers a new $4D$ gravitational theory with
only two dynamical degrees of freedom \cite{GQ4}. For example, it might
solve the singularity problem of black holes \cite{116}. Indeed, by
considering $4D$ EGB gravity in the static and spherically symmetric black
holes, Boulware and Deser \cite{G8a,G8b} obtained the first black hole
solution in the $5D$ EGB gravity, and since then steady attentions have been
devoted to black hole solutions, including their formation, stability, and
thermodynamics \cite{G9}. The shadows of spherically symmetric \cite%
{G10a,G10b}, spinning \cite{G11a,G11b}, and charged EGB black holes in AdS
space \cite{133,1334}, and other works devoted to novel $4D$ EGB black holes
have been published. However, it was shown in several papers that perhaps
the idea of the limit $D\longrightarrow 4$\ is not clearly defined, and
several ideas have been proposed to remedy this inconsistency, as well as
the absence of proper action \cite{F1,F2,F3,F4}. There are, indeed, some
correct limits or procedures, that can lead to the same black hole solutions
as naive 4D Gauss-Bonnet gravity, and different constructions were proposed,
\cite{F5,F6}. In $D\geq 5$ spacetime dimensions, BH-solutions were obtained
for vacuum \cite{G21}, anti-de Sitter (AdS) spaces \cite{G22}, and attempts
to build rotating solutions have taken place \cite{G23,G24} in the context
of Einstein Gauss-Bonnet model \cite{G25}. This paper aims to show the
influence of the Cauchy horizon by presenting the 4D EGB black hole solution
in terms of new parameterization. We use the unit system where $c=G_{N}$ $%
=4\pi \varepsilon _{0}=\hbar =k_{B}=1$.\newline
This work is organized as follows. In Section 2, we study the solution for
the charged EGB black hole in maximally symmetric vacuum and for a free
photon orbiting around along a null geodesic. Section 3, discusses the
black-hole equation of state and deals with the thermodynamic parameters,
starting with the van der Waals equation to the black hole first law. In
section 4, the discussion is extended to Extremal EGB black hole and we
introduce the compressibility for the extremal case. We conclude in the
final section.

\section{Charged Einstein-Gauss-Bonnet black hole}

\subsection{Event horizon and Cauchy horizon}

Consider now the charged Einstein-Gauss-Bonnet theory in D-dimensions with a
negative cosmological constant \cite{115,116}

\begin{equation}
I=\frac{1}{16\pi }\int d^{D}x\left( R-2\Lambda +\frac{\alpha }{D-4}G-F_{\mu
\nu }F^{\mu \nu }\right) ,  \label{EG1}
\end{equation}%
where $\alpha $ is a finite non-vanishing dimensionless Gauss-Bonnet
coupling have dimensions of $\left[ length\right] ^{2}$, that represent
ultraviolet (UV) corrections to Einstein theory, $F_{\mu \nu }=\partial
_{\mu }A_{\nu }-\partial _{\nu }A_{\mu }$ is the Maxwell tensor and $l$ is
the AdS radius and we define the Gauss-Bonnet invariant as%
\begin{equation}
G=R^{2}-4R_{\mu \nu }R^{\mu \nu }+R_{\mu \nu \rho \sigma }R^{\mu \nu \rho
\sigma },  \label{EG2}
\end{equation}%
\begin{equation}
\Lambda =-\frac{\left( D-1\right) \left( D-2\right) }{2l^{2}},  \label{EG3}
\end{equation}%
by solving the field equation we use the following spherically symmetric,%
\begin{equation}
ds^{2}=-f(r)dt^{2}+\frac{1}{f(r)}dr^{2}+r^{2}d\Omega ^{2},  \label{EG4}
\end{equation}%
where $d\Omega ^{2}=d\theta ^{2}+\sin ^{2}\theta d\phi ^{2}$ denotes the
line element of the unit $3-$sphere. Taking the limit $D\longrightarrow 4$,
we obtain the exact solution in closed form%
\begin{equation}
f(r)=-g_{00}=1+\frac{r^{2}}{2\alpha }\left( 1\pm \sqrt{1+4\alpha \left(
\frac{2M}{r^{3}}-\frac{Q^{2}}{r^{4}}-\frac{1}{l^{2}}\right) }\right) ,
\label{EG5}
\end{equation}%
where $2M$ is the Schwarzschild radius. This last solution could be obtained
directly from the derivation done in \cite{116A}. The extremal case
correspond to $f(r_{+})=0$. In the limit $\alpha \longrightarrow 0$, we can
only recover the Reissner-Nordstr\"{o}m-AdS solution. In the limit $%
r\longrightarrow \infty $ with vanishing black hole charge, we
asymptotically obtain the GR Schwarzschild solution. The event horizon in
spacetime can be located by solving the metric equation: $f(r)=0$. The
solutions show that the event horizon is located at:%
\begin{equation}
r_{\pm }=M\pm \sqrt{M^{2}-Q^{2}-\alpha }.  \label{EG8}
\end{equation}%
We notice that the solution behaves like the Reissner-Nordstr\"{o}m (RN)
solution. The black hole event horizon is the largest root of the equation
above, $r_{+}$ is the black hole horizon radius \cite{G1}. However, the
radius $r_{-}$ represents the Cauchy horizon radius. To explain the presence
of two horizon $r_{\pm }$, the mass of black holes can be rewritten as%
\begin{equation}
M^{2}=\Gamma +Q^{2}+\alpha ,  \label{EG9}
\end{equation}%
where $\Gamma =\left( r_{+}-r_{-}\right) ^{2}/2\geq 0,$ when $\Gamma =0$, we
get an extremal EGB black hole. For a non-charged EGB black hole, we find $%
M=\left( r_{+}-r_{-}\right) /\sqrt{2}$. When a black hole has a horizon $%
r_{+}>r_{-}$, the black hole is locally stable. Most papers on EGB black
hole do not indicate the importance of the Cauchy horizon $r_{-}$. Our aim
in this paper is show the influence of the Cauchy horizon radius $r_{-}$ on
the black hole thermodynamics, for that, it is necessary to study the term $%
\Gamma $.

\subsection{Maximally symmetric vacuum solutions}

We choose the gauge field, the electric potentials $\Phi _{E}$ arising from
the black hole charge $Q$, at the horizon $r_{+}$ given by
\begin{equation}
A_{\mu }dx^{\mu }=\Phi _{E}\left( r\right) dt\text{ \ \ \ and \ \ \ }\Phi
_{E}\left( r\right) =Q/r.
\end{equation}%
Note that while the scalar field possesses a harmonic time dependence, the
gauge and metric fields are static and the energy-momentum tensor will be
static and spherically symmetric. The event horizon satisfies the inequality
$r_{+}\geq M$, which implies $\Phi _{E}\left( r_{+}\right) \leq \Phi
_{E}\left( M\right) $. The cosmological constant is considered to be
dynamical, giving pressure \cite{w2}. We define the pressure \cite{127} of
the cosmological constant Eq.(\ref{EG3}) for $D\longrightarrow 4$%
\begin{equation}
P=-M_{p}^{2}\Lambda \text{ or }8\pi P=3l^{-2}\text{\ }=-\Lambda ,  \label{D0}
\end{equation}%
where $M_{p}^{2}=\frac{c^{4}}{8\pi G}\approx 2\times 10^{18}\left[ GeV\right]
\equiv 1$. In the limit of vanishing mass and charge one obtains two AdS
solutions with effective cosmological constants. The metric function can be
rewritten as%
\begin{equation}
f(\chi )=1+\chi \pm \sqrt{\left( 1+\Phi _{\Lambda }\right) \chi ^{2}-\left(
4\Phi _{G}+2\Phi _{E}^{2}\right) \chi },  \label{F}
\end{equation}%
where $\Phi _{G}=-M/r$ is the gravitational potential at a distance. On this
basis, we introduce a new order parameter\newline
$\chi (r)=r^{2}/2\alpha $ (ex: $\Gamma =\alpha \chi (r_{+}-r_{-})$), which
makes it easier to calculate the thermodynamics of black holes. We introduce
a new potential%
\begin{equation}
\Phi _{\Lambda }=\frac{4\alpha }{3}\Lambda .
\end{equation}%
When $\Phi _{\Lambda }=-1$, the vacuum energy becomes zero. Indeed, we
notice that $\Phi _{\Lambda }=-2/\chi (l)$ describes the EGB vacuum. Later
we will write all the formulas in term of $\left( \chi ,\Phi _{\Lambda
}\right) $. For AdS space, when increasing $\alpha \Lambda $ we expect that
the corresponding $\Phi _{\Lambda }$ should decrease. We notice that, in the
limit $\alpha \longrightarrow 0$ or GR limit $\left( \Phi _{G}=\Phi
_{E}=0\right) $, there exist two vacuum solutions:%
\begin{equation}
f_{-}(r)=1+\frac{\chi (r)}{\chi (l)},
\end{equation}%
\begin{equation}
f_{+}(r)=1-\frac{\chi (r)}{\chi (l)}+2\chi (r).
\end{equation}%
Thus, the $f_{-}$ branch corresponds to the standard solution, whereas the $%
f_{+}$ branch reduces to the ds/AdS with an additional term ($2\chi (r)$).
The metric reduce to the Reissner--Nordstr\"{o}m black hole solution. In
maximally symmetric vacuum solutions, there are two branches of solutions
for the effective cosmological constant \cite{116}%
\begin{equation}
\Lambda _{eff}^{\pm }=\frac{2\Lambda }{1\pm \sqrt{1+\frac{4\alpha }{3}%
\Lambda }},  \label{LL}
\end{equation}%
If $\alpha <0$\ the solution is still an AdS space, if $\alpha >0$\ the
solution is a de Sitter space, \cite{116}. By evaluating the solution Eq.(%
\ref{LL}), we get%
\begin{equation}
\Phi _{eff}^{\pm }=\frac{2\Phi _{\Lambda }}{1\pm \sqrt{1+\Phi _{\Lambda }}},
\label{L1}
\end{equation}%
these two solutions coincide when $\Phi _{\Lambda }=-1$. The potential of
the cosmological constant must obey $\Phi _{\Lambda }\geq -1$. Therefore,
the Gauss-Bonnet coupling develops a minimum $\alpha _{\min }=-3/4\Lambda $
at $\Phi _{\Lambda \min }=-1$. In the absence of a cosmological constant or $%
\Phi _{\Lambda }=0$, one obtains the event horizon of EGB black hole Eq.(\ref%
{EG8})$.$

\subsection{Free photon and null geodesics solutions}

Next, we analyze the evolution of a free photon orbiting around the
equatorial orbit of the black hole along a null geodesic, by the affine
parameter $\lambda $. The photon Lagrangian for this system can be expressed
as \cite{w3}%
\begin{equation}
\mathcal{L}=\frac{1}{2}\left[ -f(r)\dot{t}^{2}+\frac{1}{f(r)}\dot{r}%
^{2}+r^{2}\dot{\phi}^{2}\right] ,  \label{L2}
\end{equation}%
with $\dot{r}=\partial r/\partial \lambda $. We get the generalized momenta $%
p_{\mu }=\frac{\partial \mathcal{L}}{\partial \dot{x}^{\mu }}=g_{\mu \nu }%
\dot{x}^{\nu }$,\ \ \ $\mu =0,1,2,3$.\ We obtain the energy $E(=constant)$
and orbital angular momentum $L(=constant)$ of the photon, which reads%
\begin{equation}
E=-p_{t}=f(r)\dot{t}^{2}\text{ and }L=p_{\varphi }=r^{2}\dot{\varphi},
\label{L4}
\end{equation}%
with $p_{r}=\dot{r}/f(r).$The Hamiltonian the moving free photon can be
expressed as%
\begin{equation}
\mathcal{H}=2\left( p_{\mu }\dot{x}^{\mu }-\mathcal{L}\right) =-E\dot{t}+%
\frac{\dot{r}^{2}}{f(r)}+L\dot{\varphi}=0.  \label{L6}
\end{equation}%
Solving Eqs.(\ref{L4}), we easily get the equation of radial motion $\dot{r}%
^{2}=-V_{eff}$, where the effective potential in terms of $\chi $%
-parameterization is%
\begin{equation}
V_{eff}^{\pm }/V_{0}=f(\chi )/\chi -E^{2}/V_{0},  \label{L7}
\end{equation}%
where $V_{0}=L^{2}/2\alpha $. We analyse the behaviour of the effective
potential vs the parameter $\chi $ plots shown in Fig.1
\begin{figure}[H]
\centering
\includegraphics[width=16cm]{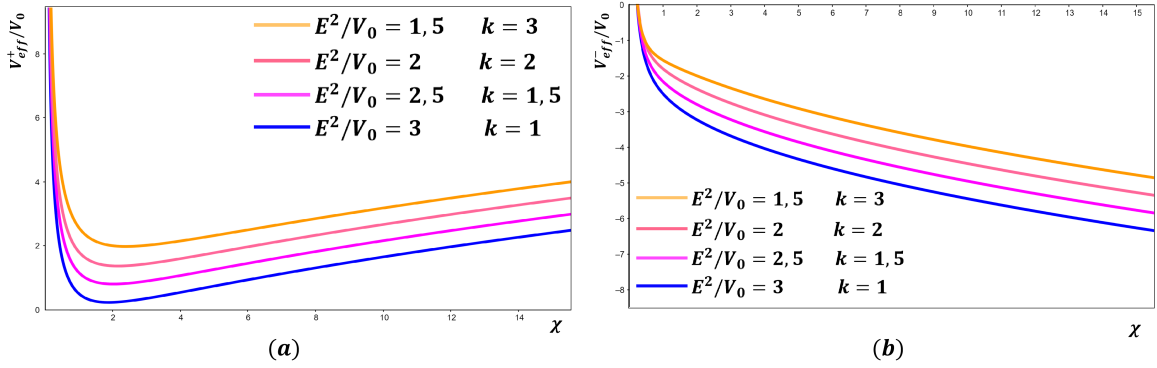}
\caption{Figure showing how the effective potential is plotted as a function
of $\protect\chi $ described by Eq.(\protect\ref{L7}). (a) and (b) denotes
the (+) and (-) part of $f(\protect\chi )$ Eq.(\protect\ref{F}),
respectively. Parameters are chosen as $\Phi _{\Lambda }=0,25$ and $k=\left(
4\Phi _{G}+2\Phi _{E}^{2}\right) $.}
\end{figure}
When the angular momentum of the free photon gets bigger in comparison to
its energy, the effects of $\alpha $ and $Q$ become weak \cite{133}. The
circular unstable photon sphere satisfies the condition
\begin{equation}
V_{eff}=\partial _{\chi }V_{eff}=0\text{ and }\partial _{\chi }^{2}V_{eff}<0,
\label{L8}
\end{equation}%
where $\partial _{\chi }=\partial /\partial \chi $. It is immediately clear
that the radius of the photon sphere $r_{ps}$ can be derived he above
conditions by solving $\partial _{\chi }V_{eff}=0$, one immediately obtains $%
2f(\chi _{ps})-\sqrt{2\alpha \chi _{ps}}\partial _{\chi }f(\chi _{ps})=0$,
where $\chi _{ps}=r_{ps}^{2}/2\alpha $. Secondly, solving $V_{eff}=0$ for a
spherically symmetric black hole, the shadow radius is $R_{S}=\sqrt{2\alpha
\chi _{ps}/f\left( \chi _{ps}\right) }$ \cite{w4}. From Eq.(\ref{F}), we can
obtain the following expression%
\begin{equation}
R_{S}=\sqrt{2\alpha }\left( \frac{1+\chi _{ps}}{\chi _{ps}}\pm \sqrt{1+\Phi
_{\Lambda }-\frac{2}{\chi _{ps}}\left( 2\Phi _{G}+\Phi _{E}^{2}\right) }%
\right) ^{-1/2},
\end{equation}%
As an example, if $\Phi _{\Lambda }=-1$ and $2\Phi _{G}=-\Phi _{E}^{2}$, in
the GR limit $\left( \alpha \rightarrow 0\right) $, the radius of the shadow
can be very well approximated by $R_{S}\sim \sqrt{2\alpha }$, that satisfies
$\chi (R_{S})=1$. Within this limit, the surface of the shadow becomes
\begin{equation}
A_{S}\sim 8\pi \alpha .  \label{AA}
\end{equation}%
To reach a maximum value of the shadow surface, using Eq.(\ref{EG8}), we
find $A_{S\max }=8\pi \left( M^{2}-Q^{2}\right) $, which means that for $M=Q$%
, we obtain a black hole without shadow. In the following, we would like to
express logarithmic term of EGB entropy by $A_{S}$.

\section{Thermodynamics of 4D EGB Black hole}

\subsection{The van der Waals equation}

We first compute the mass, temperature and entropy of the EGB black hole
\cite{JEP}, in order to analyse the validity of the second law of black hole
thermodynamics. We can express the ADM mass $M$ of the black hole in terms
of $r_{+}$, by solving $f(r)=0$ for $r=r_{+}$ resulting in%
\begin{equation}
M=\frac{l^{-2}r_{+}^{4}+r_{+}^{2}+Q^{2}+\alpha }{2r_{+}}.  \label{M0}
\end{equation}%
The temperature of the black hole corresponds to the tangent of the Newton's
potential at the event horizon \cite{JEP}.The Hawking temperature $T=\beta
^{-1}$ of the black hole can be calculated as \cite{116}

\begin{equation}
T=\frac{1}{4\pi }f^{\prime }(r_{+})=\frac{3l^{-2}r_{+}^{4}+r_{+}^{2}-Q^{2}-%
\alpha }{8\pi \alpha r_{+}++4\pi r_{+}^{3}}.  \label{T0}
\end{equation}%
\newline
One recovers the temperature of the Schwarzschild black hole in the limit of
$\left( \alpha =0,Q=0\text{ and }l^{-2}=0\right) $, in which cas $T=1/4\pi
r_{+}$ as expected. For $Q$ $>0$, $T>T_{Sch}$, where $T_{Sch}$ is the
temperature of Schwarzschild black hole. On the contrary, at $Q<0$, $%
T<T_{Sch}$. We can express the Eq.(\ref{T0}) as
\begin{equation}
\frac{1+\chi _{+}}{\upsilon \chi _{+}}T=\left( P+\frac{1}{16\pi \alpha \chi
_{+}}\right) +\frac{\Gamma -M^{2}}{4\pi \alpha \chi _{+}\upsilon ^{2}},
\label{T2}
\end{equation}%
where $\chi _{+}=\chi (r_{+})=A/8\pi \alpha $. We introduce the
thermodynamic volume $V=\left( \partial M/\partial P\right) _{S,Q,...}=4\pi
r_{+}^{3}/3$. We use the specific volume $\upsilon =2r_{+}l_{P}^{2}\equiv
2r_{+}=6V/N>$ $0$, where $l_{P}=\sqrt{\hbar G/c^{3}}\equiv 1$ is the Planck
length and $N=A/l_{P}^{2}$ is the number of states associated to the horizon
\cite{132}. Considering the transformations \cite{w7}, the van der Waals
equationcan be obtained as%
\begin{equation}
P_{\chi }=P+\frac{1}{16\pi \alpha \chi _{+}}\text{, \ }a_{\chi }=\frac{%
M^{2}-\Gamma }{4\pi \alpha \chi _{+}}\text{ and\ \ }b_{\chi }=\frac{\upsilon
}{1+\chi _{+}}.  \label{T3}
\end{equation}%
The change of the black hole parameters are $\left( P_{\chi },\text{ }%
a_{\chi },\text{ }b_{\chi }\right) $. In this case, the equation of state
takes the form%
\begin{equation}
\left( P_{\chi }+\frac{a_{\chi }}{\upsilon ^{2}}\right) \left( \upsilon
-b_{\chi }\right) =T\text{,}  \label{T4}
\end{equation}%
where the parameter $a_{\chi }$ is the average attraction, which measures
the attraction between particles. The parameter $b_{\chi }$ corresponds to
the volume of fluid of particles. From the last equation, by taking $\Gamma
=M^{2}$ and $\chi _{+}\sim 0$, i.e. non-charged EGB black hole, we can
easily recover the ideal gas equation $P_{\chi }\upsilon =T$. We notice that
there is an appearance of term $\Gamma $ in $a_{\chi }$. If $\Gamma $ is
very low, the attraction between particles will be very large. One concludes
that $\Gamma $ maps the interacting gas into an ideal gas of point
particles. The attraction corresponds to $a_{\chi }>$ $0$ i.e. $\Gamma <M^{2}
$. To show the effects of the temperature on $P_{\chi }-\upsilon $ isotherms
of the van der Waals equation of state, we have plotted Fig.2.
\begin{figure}[H]
\centering
\includegraphics[width=16cm]{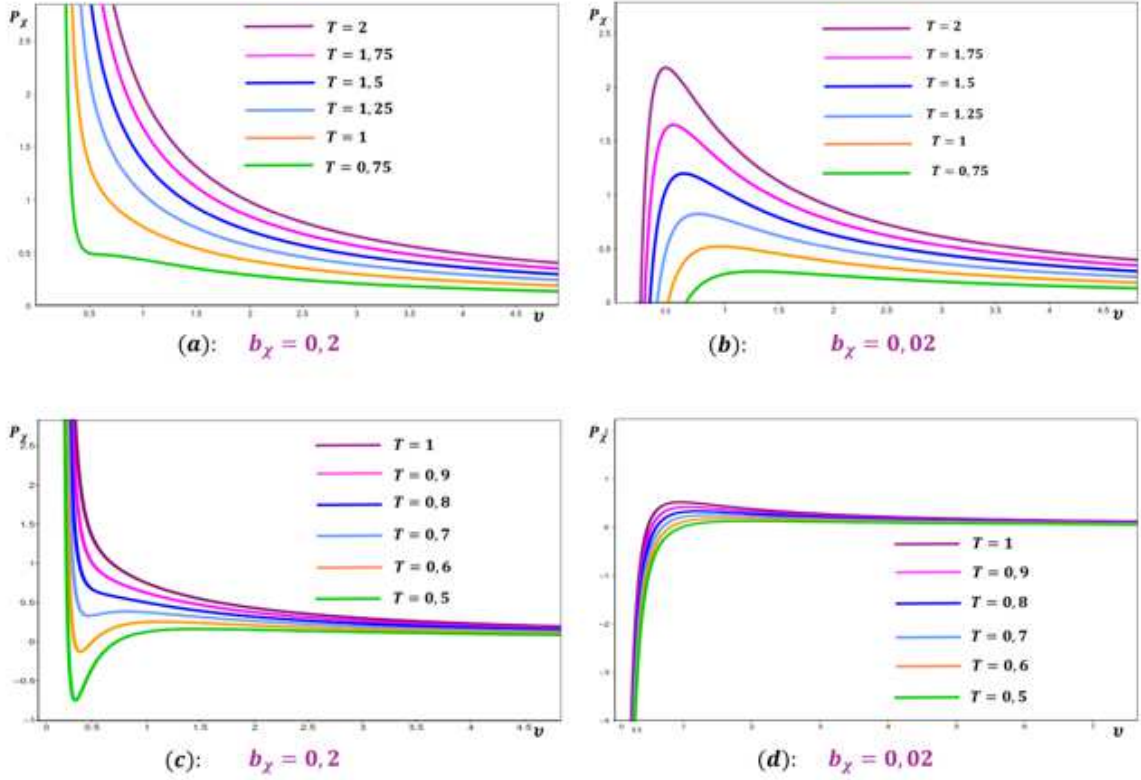}
\caption{$P_{\protect\chi }-\protect\upsilon $ isotherms of the van der
Waals equation of state, with fixed parameters $a_{\protect\chi }=0,5$ and $%
b_{\protect\chi }=0,2$ (left) $b_{\protect\chi }=0,02$ (right).}
\end{figure}
For the choice of the transformations%
\begin{equation}
P_{\chi }=P+\frac{1}{16\pi \alpha \chi _{+}}\text{, \ }a_{\chi }=\frac{%
M^{2}-\Gamma }{4\pi \alpha \chi _{+}}\text{ and\ }T_{\chi }=\frac{1+\chi _{+}%
}{\chi _{+}}T,  \label{T5}
\end{equation}%
we get the equation of state%
\begin{equation}
\left( P_{\chi }+\frac{a_{\chi }}{\upsilon ^{2}}\right) \upsilon =T_{\chi }.
\label{T6}
\end{equation}%
It is interesting that the existence of a first-order phase transition
between small and large black hole corresponding to a liquid and gas phase
transition that automatically terminates at a critical point $\left(
P_{c},v_{c},T_{c}\right) $ eventually satisfied $\left( \partial _{\upsilon
}P_{\chi }\right) _{T...}=\left( \partial _{\upsilon }^{2}P_{\chi }\right)
_{T...}=0$ \cite{w7}.
\begin{figure}[H]
\centering
\includegraphics[width=10cm]{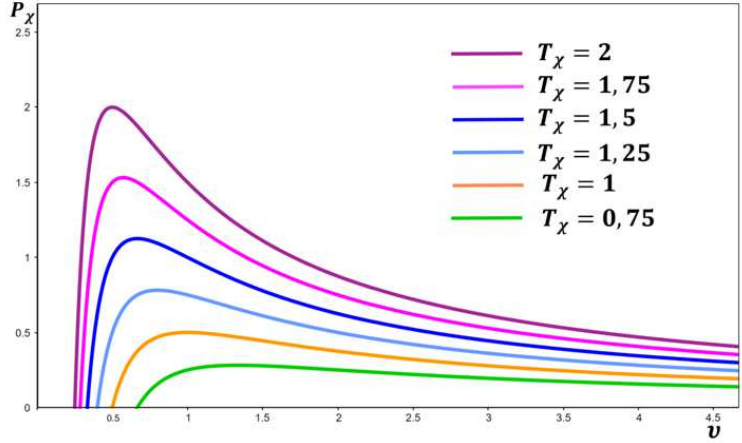}
\caption{$P_{\protect\chi }-\protect\upsilon $ isotherms of the van der
Waals equation of state.}
\end{figure}
We shall now describe the Joule-Thomson effect, which is entirely related to
the difference between a real gas and an ideal gas, especially the
attraction and repulsion of the van der Waals forces. The Joule-Thomson
coefficient $\mu $ is defined as%
\begin{equation}
\mu =\left( \frac{\partial T}{\partial P}\right) _{M}.  \label{T7}
\end{equation}%
For an ideal gas, $\mu $ is always equal to zero. The Joule-Thomson
inversion temperature is the temperature for which the coefficient $\mu $
changes sign. It is easy to see that the system will experience a cooling
(heating) process with $\mu $ $>$ $0$ $(\mu <0)$, caused by the change in
pressure is always negative during expansion. Let us consider the case Eq.(%
\ref{T4}):%
\begin{equation}
\mu _{\chi }=\frac{\chi _{+}}{1+\chi _{+}}\upsilon .  \label{T8}
\end{equation}%
If $\mu _{\chi }>$ $0$, which means $\chi _{+}>$ $0$ or $\chi _{+}<$ $-1$,
then the gas temperature is below the inversion temperature. The interval $%
-1<\chi _{+}<0$, implies that $\mu _{\chi }<$ $0$, then the gas temperature
is above the inversion temperature. If we keep the specific volume fixed and
vary the coefficient $\chi _{+}$ near $-1$or $0$, then there is a inversion
temperature similar to that of the phase transition. It can be explicitly
verified that the inversion temperature corresponding to the evolution of
the parameter $\chi _{+}$. Consider, for instance Eq.(\ref{T6}), it is
immediately clear that $\mu _{\chi }=\upsilon >0$. Then the gas temperature
is above the inversion temperature. The important issue that requires
attention is the role of $\chi _{+}$ appears in Eq.(\ref{T4}), and not in
Eq.(\ref{T6}). For that, in what follows we will study only the case Eq.(\ref%
{T4}). To study the effect of $\chi _{+}$, we have plotted Fig.4.
\begin{figure}[H]
\centering
\includegraphics[width=9cm]{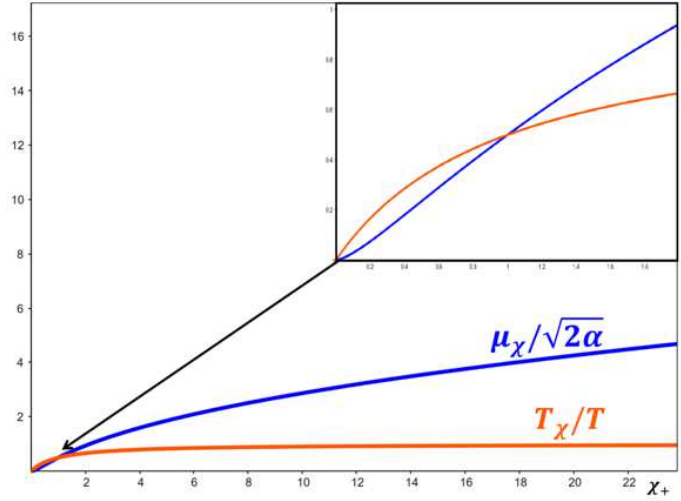}
\caption{$\protect\mu _{\protect\chi }/\protect\sqrt{2\protect\alpha }$ and $%
T_{\protect\chi }/T$ vs $\protect\chi _{+}$.}
\end{figure}
To develop the average attraction $a_{\chi }$ of the new parametric
expressions $\left( \chi _{+},\Phi _{\Lambda }\right) $, it is helpful to
first review the evolution of the thermodynamic variables, based on Ref.\cite%
{w2}.
\begin{equation}
a_{\chi }/4=\left( P_{\chi }+\frac{3}{4}\frac{\Phi _{\Lambda }}{8\pi \alpha }%
\right) \left( M^{2}-\Gamma \right) ,  \label{T9}
\end{equation}%
this result is similar to that of the van der Waals equation. We imagine a
scenario where the equation above represents a van der Waals equation. For
this, one must have $8\pi \alpha \sim M^{2}$. Then we can impose a new
temperature $T_{a}\propto a_{\chi }/4$, which leaves only. Two-parameter
equation of state induced by the transformations Eq.(\ref{T3}) of Ref.\cite%
{w2}, which is the expression of van der Waals equation for the temperature $%
T_{a}$. The term $M^{2}$ can be shown to be a specific volume and $\Gamma $
can be a new volume excluded. Whereas, in asymptotic limit $4P_{\chi }\ll
\frac{3\Phi _{\Lambda }}{8\pi \alpha }$, it is then clear that%
\begin{equation}
\Gamma \sim M^{2}-\frac{a_{\chi }}{4P_{\chi }}.  \label{gama}
\end{equation}%
\emph{Solution without particles}: First, let us investigate the ADM mass
Eq.(\ref{M0}) given by%
\begin{equation}
\Gamma =\left( M-\frac{\upsilon }{2}\right) ^{2}-\frac{NP\upsilon ^{2}}{6}.
\label{ma}
\end{equation}%
Indeed, if we compute $N=0$ (black hole without particles), in the limit $%
\upsilon \ll 2M$, the ideal gas equation can be done by comparing Eqs.(\ref%
{gama},\ref{ma}), at this point, we get $T_{ideal}\equiv a_{\chi }/4M\sim
P_{\chi }\upsilon $. These results can be generalized to $N\neq 0$ and
without asymptotic limit. Consequently, Eq.(\ref{T9}) take the form of the
van der Waals equation. This analysis, leads us to find, $T_{a}\equiv
a_{\chi }/4M$.

\subsection{The black hole first law}

For EGB-AdS black hole, the mass of a black hole is more appropriately
interpreted as enthalpy $H$. The black hole first law reads \cite{127}%
\begin{equation}
dM=\Phi _{E}dQ+Ad\alpha +TdS+VdP,  \label{D1}
\end{equation}%
where $A\equiv 4\pi r_{+}^{2}$ is the area of the event horizon of the black
hole. The parameters $V$ and $A$ are the conjugate quantities of the
pressure $P$ and GB coupling parameter $\alpha $, respectively \cite{G5}. If
we fix $P$ and $\alpha $ Eq.(\ref{D1}), the Bekenstein-Hawking formula is
then given by%
\begin{equation}
S=\int \frac{dM}{T}=\frac{A}{4}+2\pi \alpha \log \frac{A}{A_{0}},  \label{EN}
\end{equation}%
where $A_{0}$ is some constant with units of area. An important
justification for this new logarithmic term in Eq.(\ref{EN}), came from the
Eq.(\ref{AA}), on using shadow surface, yields $2\pi \alpha =A_{S}/4$, which
means that the logarithmic term (added by EGB gravity) depends on the shadow
of black hole. We can express the EGB black hole entropy in terms of the
shadow surface, one gets
\begin{equation}
S=\frac{A}{4}+\frac{A_{S}}{4}\log \frac{A}{A_{0}}.
\end{equation}%
This investigation reveals a potential relationship between the entropy and
the black hole shadow. These results can be generalized to the entropy of
the 4D Hayward-EGB black hole \cite{w1}. Further in the GR limit $\left(
\alpha \rightarrow 0\right) $, we get the entropy obeying area law (ex: the
Schwarzschild black hole) \cite{w6}, the same if $A\rightarrow A_{0}$. Eq.(%
\ref{EG9}) is based on the fact that, as thermodynamic systems, black holes
must obey the first law of thermodynamics%
\begin{equation}
dM=\frac{Q}{M}dQ+\frac{1}{2M}d\alpha +\frac{1}{2M}d\Gamma ,  \label{D2}
\end{equation}%
where $\Phi _{E}\left( r_{+}\right) \leq \Phi _{E}\left( M\right) $. We now
turn to determining the the entropy of the resulting black hole by another
method. This can be done by comparing Eqs.(\ref{D1},\ref{D2}), we find
\begin{equation}
dS=\frac{d\Gamma }{2MT},  \label{wc}
\end{equation}%
which results into the following relationship (\ref{gama}):%
\begin{equation}
\frac{d\Gamma }{2MT}\sim \frac{dM}{T}-\frac{da_{\chi }}{8P_{\chi }MT}+\frac{%
a_{\chi }dP_{\chi }}{8P_{\chi }^{2}MT},
\end{equation}%
which turns into the expression of the entropy of 4D EGB black hole Eq.(\ref%
{EN}), when we fix $P_{\chi }$ and $a_{\chi }$ Eq.(\ref{T3}): $S\sim \int
dM/T$. This means that $\Gamma $ is a generator of the black hole entropy.
Therefore, since $\Gamma $ depends on the Cauchy horizon radius, this shows
that $r_{-}$ has a great effect on the thermodynamic parameters.

\section{Extremal EGB black hole}

We now describe as examples, the extremal EGB black hole. If we fix $P$ and $%
\alpha $ and by comparing Eq.(\ref{D1}) with Eq.(\ref{D2}), we indicate that%
\begin{equation}
\Phi _{E}=\frac{Q}{M},\text{ \ }\frac{d\Gamma }{2M}=TdS-PdV.  \label{D3}
\end{equation}%
Next, we consider that $\alpha $ is fixed. As is well known that, the
electric permittivity $\epsilon _{S}=\left( \partial Q/\partial \Phi
_{E}\right) _{S}=M$ \ is relevant for stability in the grand canonical
ensemble. Regarding this eqaution, we get an extremal black hole solution
with degenerate horizon given by \qquad
\begin{equation}
r_{+}=r_{-}=M=\sqrt{Q^{2}+\alpha }.  \label{M}
\end{equation}%
In this case, the Hawking temperature reduces to\ \
\begin{equation}
T=\frac{2Pr_{+}^{3}}{2\alpha +r_{+}^{2}}  \label{D5}
\end{equation}%
The entropy for this black hole can be obtained as%
\begin{equation}
S=\int \frac{P}{T}dV=\frac{A}{4}+2\pi \alpha \log \frac{A}{A_{0}}\sim 2\pi
\alpha \log \left[ \chi _{+}\frac{1+\chi _{+}}{\chi _{0}}\right] ,
\label{D6}
\end{equation}%
where $A_{0}$and $\chi _{0}$ are some constants with units of area. This
expression coincides with the entropy formula in \cite{G5} (we have showed
this entropy by a different method). $P$ and $\alpha $ are constant for an
extremal black hole. This entropy led the study of black holes as a
thermodynamical system. From Eq.(\ref{D1}), we get the equation of state for
the extremal EGB black hole as follows%
\begin{equation}
P\upsilon =\frac{1+\chi _{+}}{\chi _{+}}T,  \label{D7}
\end{equation}%
where $\chi _{+}=\chi (r_{+})\neq 0$. The extremal EGB black hole behaves
like an ideal gas if $\chi _{+}=c$. To determine the thermodynamic
similarity class of different substances we use compressibility factor $%
Z=P\upsilon /T>0$ \cite{BH4} in terms of the specific volume%
\begin{equation}
Z=1+\frac{1}{\chi _{+}},  \label{D8}
\end{equation}%
we analyze in detail the thermodynamics and phase transitions for exact. The
black holes \emph{with} $\chi _{+}<0$, the compressibility factor is $Z<1$,
this shows that there is a great interaction between the gas particles. The
pressures are lower, the particles are free to move. In this case, the
attractive forces dominate. \emph{In GR limits} ($\alpha =0$): $\chi
_{+}\rightarrow \infty $, the compressibility factor is $Z=1$, this shows
that the gas behaves like an ideal gas. In this case, Eq.(\ref{D7})
represents the equation of state in the horizon. \emph{For} $\chi _{+}>0$,
the compressibility factor is $Z>1$, the particles collide more often. This
allows the repulsive forces between molecules to have a noticeable effect.
It is then clear that the relation between $Z$ and $\alpha $ is expressed as
$Z-1=2\alpha /r_{+}^{2}$, this means that the parameter $\alpha $ increased
the repulsion between the particles. Therefore, this $\alpha $ maps the
ideal gas into a noninteracting gas. From Eqs.(\ref{D5}), we calculate the
specific heat of 4D extremal EGB black hole by%
\begin{equation*}
C=\frac{\partial M}{\partial T}=\frac{3\Phi _{\Lambda }}{4\pi T}\frac{\chi
_{+}^{2}}{3+\chi _{+}}.
\end{equation*}%
The thermodynamical systems is the locally stable if $C>0$ (or $\chi _{+}>-3$%
), and is unstable if $C<0$ (or $\chi _{+}<-3$). Making use of the explicit
case $\chi _{+}>-3$, signifies that the extremal black holes are
thermodynamically stable against perturbations in the region. For a
thermodynamical systems, The behaviour of specific heat leads to find the
regions of local and global stability of the 4D extremal EGB black hole. The
diverging specific heat from $C<0$ to $C>0$, implies the existence of
second-order phase transition \cite{w1}.

\section{Conclusion}

First, we have considered the four-dimensional charged Einstein-Gauss-Bonnet
theory. We also showed that the thermodynamic processes of black holes in
AdS, can be modeled by the parameter $\chi _{+}$. The Joule-Thomson effect
is entirely related to the difference between a real and an ideal gas,
especially the attraction and repulsion of the van der Waals forces. To
study the influence of the Cauchy horizon radius, we have introduced the
parameter $\Gamma $, which depends at the same time on the Cauchy horizon
radius and the event horizon radius. In other words, the term $\Gamma $ maps
the interacting gas into an ideal gas of point particles. It makes the
attraction decrease between particles. The results showed that all the
thermodynamic quantities depend on the GB coupling parameter. The
coexistence curve in the $P_{\chi }-\upsilon $ diagram was shown, which is
similar to the van der Waals fluid. After considering the GR limit for the
evolution of a free photon orbiting along a null geodesic, we found the
shadow surface. They satisfy the simple relation $A_{S}\sim 8\pi \alpha $.
Hence, we have obtained a relationship between the shadow surface and the
black hole entropy. Finally, we investigated 4D extremal EGB black hole as a
working substance and studied the entropy, the equation of state,
compressibility and the diverging specific heat.

\end{document}